# Self-pulsation in fiber-coupled on-chip microcavity lasers

Lina He, Sahin Kaya Ozdemir, Jiangang Zhu, and Lan Yang

Department of Electrical and Systems Engineering, Washington University in St. Louis, St. Louis, MO 63130, USA

Corresponding author: yang@ese.wustl.edu, ozdemir@ese.wustl.edu

**Abstract** We report self-pulsation in an erbium-doped silica toroidal microcavity laser coupled to a tapered fiber and investigate the effects of pump power and taper-cavity coupling condition on the dynamic behaviors of the pulse train. The microcavity is pumped at 1444.8 nm and lasing occurs at 1560.2 nm with a threshold of 12 μW. Experimental results are interpreted within the framework of ion-pair induced self-quenching model.

OCIS: 230.5750, 140.3538, 140.3500, 220.4000

Recent developments in on-chip optical microcavities with high-quality factors and microscale mode volumes open a promising direction to realize low-threshold lasers which can be easily integrated with silicon-based photonic components. Low-threshold lasing in microresonators has been demonstrated by either taking the advantage of high power build-up in high-$Q$ microcavities, e.g., microcavity Raman lasers [1], or integrating active materials into cavity structures, e.g., rare-earth ions [2] or quantum dots [3] doped microcavity lasers. Rare-earth ions doped microcavity lasers have attracted much interest due to their easy incorporation into host materials, high efficiencies, long excited-state lifetimes and widely-spanning emission spectra which cover wavelengths crucial to many sensing and communication applications, e.g., Erbium (Er) provides gain around 1550 nm which makes it a key dopant for optical communication applications. Lasing from Er-doped microcavities have been demonstrated by either fabricating the microcavities from Er-doped glass [4, 5], or coating the microcavities with Er-doped sol-gel films [2, 6].

Yang *et al.* reported that Er-doped microcavities can sustain continuous-wave (CW) and self-pulsing (SP) operations at low and high ion concentrations, respectively [2]. Similar operations were observed and studied in rare-earth-ion doped optical fiber lasers and amplifiers [7, 8]. Self-pulsing in heavily Er-doped laser systems under continuous pumping is attributed to ion-ion interactions in Er clusters leading to cooperative energy transfer between ions which returns one ion to ground state preventing the population of paired ions from full inversion [8]. Lasers based on high-$Q$ whispering gallery modes (WGMs) microcavities hold advantages of lower threshold, compact size, and easy integration with on-chip devices. In addition, microcavity laser excited through a fiber taper provides an easy way to tune the cavity lifetime as well as the laser dynamics via adjusting the taper-cavity air gap. Previously, Min *et al.* studied CW operation in Er-implanted toroidal microcavity lasers including the effects of cavity loading and Er concentration on the lasing wavelength and threshold [9]. To date, factors affecting the dynamics of SP in Er-doped microcavities have remained as open problems. In this Letter, we demonstrate and systematically investigate SP in Er-doped silica microtoroidal lasers with special attention to the effects of pump power and cavity loading condition on the time evolution of laser pulses.

Microcavities used in this work are fabricated from Er-doped sol-gel silica layer on a silicon wafer through photolithography procedure followed by $CO_2$ reflow [5]. Results reported in this Letter are obtained from a microtoroid with major (minor) diameter of 43 μm (5 μm) and Er concentration of $2\times10^{19}$ ions/cm$^3$. Experimental setup used to characterize the Er-doped microtoroid laser is shown in Fig. 1. A tunable external cavity CW laser provides the pump light in 1460 nm wavelength band. Triangle wave from a function generator is utilized to perform fine scan of the pump wavelength around a high-$Q$ WGM of the microcavity within the absorption band of Er ions. A polarization controller is used to adjust the polarization state of pump light to maximize the pump-cavity coupling efficiencies, and therefore achieve lasing at a lower threshold power in the fiber. Pump light is coupled into the Er-doped microtoroid via a tapered fiber with a waist diameter of 1~2 μm [10]. Cavity loading condition is adjusted by varying the air gap between the taper and the toroid through a nano-positioning system. The same taper is used to retrieve laser light from the microtoroid. Output of the cavity is connected to a 10/90 fiber coupler, with the 10% port connected to an optical spectrum analyzer, while the 90% port directed to a fiber based wideband wavelength division multiplexer (WDM) to separate the laser emission from the unabsorbed pump light. Both outputs of the WDM are monitored by 125 MHz photodetectors whose electrical signals are sent to an oscilloscope connected to a computer for data acquisition and processing.

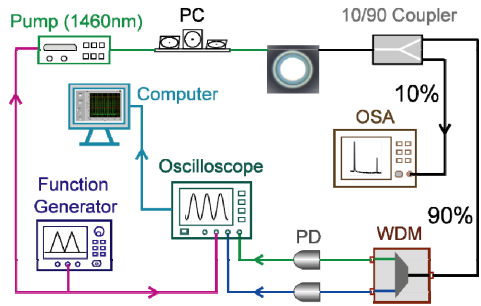

Fig. 1. Experimental setup. PC: polarization controller, OSA: optical spectrum analyzer, WDM: wavelength division multiplexer, PD: photodetector.

Figure 2 shows a typical single mode emission spectrum and pulse train of an Er-doped microtoroid excited above threshold. Laser emission is obtained at $\lambda_s$=1560.2 nm for the pump at $\lambda_p$ =1444.8 nm. As expected there is a clear linear dependence of the output laser power on the input pump power above the threshold 12 μW. The slope efficiency is calculated as 2%. Particularly interesting observation is the SP due to saturable absorbers formed by cross-relaxation process between paired ions. Contrary to other forms of oscillations which are observed in high-$Q$ optical microcavities (e.g., mechanical vibrations due to radiation pressure [11] and oscillations due to beating of counter-propagating [12] or co-propagating [13] WGMs), SP oscillations have a unique non-sinusoidal waveform. The fluctuations of lasing intensity might originate from thermal effects [14], mechanical oscillations [11], the fact that population inversion density may not return to the same value each time the cavity dumps out the pulse [15] or perturbations in pump power. For experiments reported in this Letter, the Er-doped microlaser is operated at the same single mode regardless of the taper-cavity air gap [9].

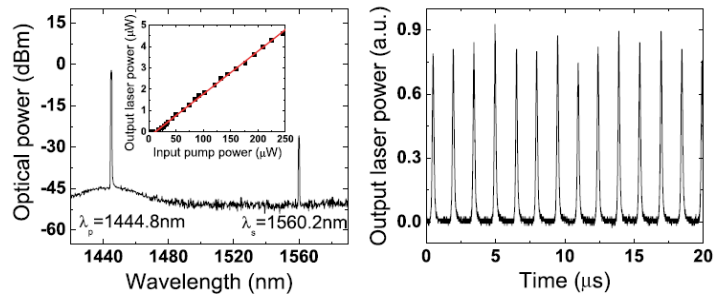

Fig. 2. Typical single mode laser spectrum and pulse train from an Er-doped sol-gel silica microtoroid laser. Inset shows the measured output laser power as a function of the launched pump power into the fiber.

To study the effect of pump power on the laser pulses, we monitor the changes in the pulse train in response to variation in the launched pump power. The taper-cavity air gap is set close to the critical coupling point of the pump mode to maximize the pump power coupled into the cavity. Figure 3 shows that the measured pulse repetition time ($T$) and full-width at half-maximum ($\Delta\tau$) decrease with increasing pump power. The results are similar to those observed in Er-doped fiber lasers and can be explained by the fact that pumping the gain to its threshold to overcome the total losses (saturable and unsaturable losses) and bleaching the saturable absorbers are achieved much faster at higher pump powers [7,8]. Thus, time interval between the end of a pulse and the start of the consecutive pulse becomes shorter for higher pumping rates. Rise time of the pulse depends on the increase rate of the net gain which is affected by pumping rate and cavity lifetime $\tau_c$. Fall time is related to the recovery time of saturable absorbers and the rise time, given that the rise time is much longer than $\tau_c$ in our experiments [15, 16]. Since the cavity loading condition is kept constant for the measurements in Fig. 3, pump power plays the main role in determining $T$ and $\Delta\tau$.

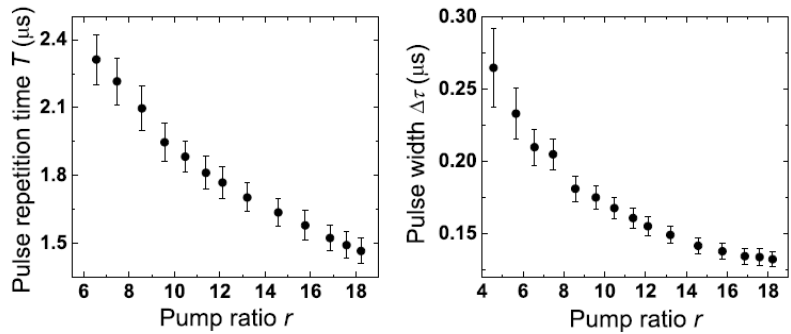

Fig. 3. Dependence of pulse repetition time and pulse width on pump ratio $r = P_{in}/P_{th}$ where $P_{in}$ and $P_{th}$ are the input and threshold pump power, respectively. Vertical bars denote the standard deviations.

Loading condition of the taper-cavity system is crucial in determining the SP dynamics, as it affects (i) pump-cavity coupling efficiency and hence the intra-cavity pump power, and (ii) coupling loss of the lasing mode which in turn determines the lasing threshold, extraction of laser power from the microcavity, and cavity lifetime [17]. To investigate the effects of cavity loading condition on the SP dynamics, we set the input power to the fiber well above lasing threshold and record the SP operation in under-, critical- and over-coupling regimes while the taper-cavity air gap is finely tuned. Figures 4 and 5 show how the interplay of (i) and (ii) affects the peak power $P_{peak}$, $T$ and $\Delta\tau$ of the laser pulses.

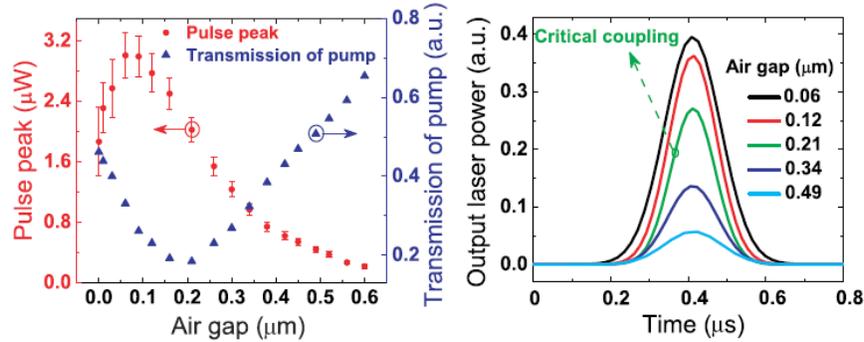

Fig. 4. Output laser power versus air gap (left) and Gaussian fit of single pulse at various air gaps (right).

Starting from deep under-coupling regime, as the air gap decreases, $P_{peak}$ increases gradually until it reaches its maximum $P_{max}$ in the over-coupling regime (air gap < 0.2 µm) of the pump-cavity loading curve rather than at the critical coupling point (air gap ~ 0.2 μm) where the intra-cavity pump field reaches its maximum value (Fig. 4). Further decrease of the air gap reduces $P_{peak}$. Factors affecting the position of $P_{max}$ are explained as follows. (i) Taper-cavity coupling for the lasing mode: This determines the extraction rate of the emitted laser photons from the cavity, and favors the over-coupling regime in which stronger taper-lasing mode coupling allows extracting higher energy from the active medium in a specified time interval. (ii) Amount of energy stored in the active medium: This is determined by the saturable absorber loss, cavity lifetime for the lasing mode and the pumping rate. Smaller cavity lifetime in the over-coupling regime leads to a slower build-up rate of the laser pulse inside the cavity. Together with the saturable absorber loss and the continuous pumping, it allows the gain building up to a higher value giving rise to higher energy storage in the active medium. As a result, the optimum coupling condition for maximum laser output power is pushed towards the over-coupling regime, where the laser photons in the resonator experience higher coupling to the fiber [16]. (iii) Taper-lasing/pump mode phase-mismatch: This is determined by the difference between the propagation constants of the lasing/pump mode in the microtoroid and the taper mode. While the former one is fixed for the studied lasing/pump mode in the microcavity, the latter can be adjusted by the taper size. Increased phase-mismatch leads to changes in coupling coefficients for both lasing and pump modes which in turn affect the position of $P_{max}$. In our experiments, we observed that the location of $P_{max}$ favored over-coupling regime of the pump-cavity coupling due to the interplay of (i)-(iii).

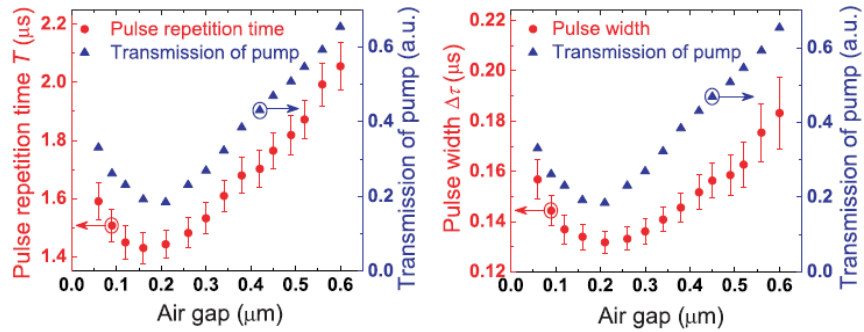

Fig. 5. Dependence of pulse repetition time and pulse width on air gap.

Figure 5 shows the effect of air gap, i.e., taper-cavity coupling, on $T$ and $\Delta\tau$. Starting from the under-coupling regime, both $T$ and $\Delta\tau$ monotonously decrease until critical coupling point where they attain their minimum values. By further decrease of air gap into over-coupling regime, both $T$ and $\Delta\tau$ show increasing trends. These observations are consistent with those presented in Fig. 3 as more pump power is coupled into the cavity when the taper-cavity system gets closer to the critical coupling point [17]. For identical pump transmission in under- and over-coupling regimes, $\Delta\tau$ is always longer in over-coupling regime while $T$ has similar values within the standard deviations in both regimes. The longer $\Delta\tau$ can be explained as follows: Laser photons generated from stimulated emission of excited Er ions are extracted from the cavity at a higher rate in over-coupling regime due to the stronger taper-cavity coupling. The resulted slower build-up rate of laser pulse within the cavity slows down the dumping of the population inversion thus leading to longer $\Delta\tau$. On the other hand, $T$ strongly depends on the pump power and does not show significant observable variation within its time scale because intra-cavity pump powers are equal in under- and over-coupling regimes for identical values of transmitted pump power.

In conclusion, we have demonstrated and studied self-pulsing in taper-coupled toroidal microcavity lasers fabricated from Er-doped sol-gel silica film. The SP operation was characterized by the peak power of the pulse train, pulse repetition time and pulse width. Dependence of SP dynamics on pump power and taper-cavity coupling condition was investigated experimentally. The optimal extraction of laser light from the cavity does not coincide with the critical coupling point for the pump mode. This manifests itself as a shift of optimum coupling for maximum laser output power from the critical coupling point in pump-cavity loading curve. Our results also show that the pulse repetition time and pulse width can be tuned by the taper-cavity air gap.

This work is supported by the National Science Foundation under Grant No. DMR0907467.